\documentclass[a4paper,10pt]{article}
\usepackage{multicol}
\usepackage{amsmath}
\usepackage{amssymb}
\usepackage{graphicx}
\usepackage{float}
\usepackage{amsmath}
\usepackage{bm}
\usepackage{amsfonts}
\usepackage{balance}
\usepackage{titling}
\makeatletter
\def\@cite#1#2{{#1\if@tempswa,#2\fi}}
\def\@biblabel#1{\textsuperscript{#1}}
\newcommand\figcaption{\def\@captype{figure}\caption}
\makeatother
\newcommand{\ct}[1]{{\textsuperscript{{\cite{#1}}}}}
\newcommand{\bee}{\begin{equation}}
\newcommand{\ee}{\end{equation}}
\newcommand{\beea}{\begin{eqnarray}}
\newcommand{\eea}{\end{eqnarray}}

\newcommand{\vv}[1]{{\mathbf #1}}

\newcommand{\ci}{\textrm{i}}

 \newlength{\halfpagewidth}
 \divide\halfpagewidth by 2

\thanksfootextra{}{)}
\thanksheadextra{}{)}
\thanksmarkseries{alph}
\begin{document}
\title{Faraday rotation effect in periodic graphene structure}
\author{ {Daqing Liu,$^{1}$\thanks{liudq@mail.xjtu.edu.cn} ~~Shengli Zhang,$^{1}$\thanks{zhangsl@mail.xjtu.edu.cn} ~~Ning Ma,$^{1}$\thanks{maning@stu.xjtu.edu.cn} ~~and Xinghua Li$^{2}$\thanks{lixh@mail.fjnu.edu.cn}} \\
{\small \it $^{1}$ Department of Applied Physics, Non-equilibrium Condensed Matter and Quantum
        Engineering Laboratory, }\\
        {\small \it KLME, Xi'an Jiaotong University, Xi'an 710049, China }\\
{\small \it $^{2}$ Physics and Optic-Electronics Technology College, Fujian Normal University, Fuzhou 350007, China}
 }
\date{}
\maketitle
\begin{minipage}{0.85\textwidth}
We report the magneto-optical rotation effect in a periodic graphene-sheet structure. Due to the masslessness of carriers in graphene, the magnetic response is very sensitive and the magneto-optical rotation effect is therefore significant. We predict that the Verdet constant of the periodic graphene-sheet structure is roughly $10-100$ times that of rare-earth-doped magneto-optical glass in the infrared region.


\end{minipage}

\vskip 0cm

\begin{multicols}{2}

\section{INTRODUCTION}

In the early 1960s,
magneto-optical (MO) properties\ct{faraday} of materials
started to develop rapidly because of the work of Bell Laboratories. Theoretical improvement
and a lot of MO material synthesis led to developing many
MO devices, such as MO
modulators, MO isolators, MO sensors,
magnetic optical circulators and MO memory, and even
magnetometry.\ct{metry} However, rotatory power is not very
high in usual materials. For instance, even for typical high-Verdet
materials, such as MO glasses, the Verdet constant (VC) is only on
the order of $0.1-1\,\mathrm{min/(Oe\cdot cm)}$.\ct{high} Realizing
a large VC is still a theoretical and experimental challenge.

Surprisingly, Ref. \cite{nature} finds that we can obtain several degrees of rotation of the polarized direction when a linear electromagnetic wave passes through a single atomic layer of carbon, graphene. Recalling that graphene is the ultimately thin material in condensed matter physics, we regard this as a significant effect. There are also other studies\ct{prb1,apl,prb2,prl2,phil} on this topic. For instance, Ref. \cite{prb1} studies the phenomenon using the equation of motion, and Ref. \cite{apl} shows that graphene exhibits extremely broadband nonreciprocal polarization rotation at subterahertz frequencies.

So far studies have focused on rotation in monolayer, bilayer or multilayer graphene.
To make a practical MO device, that is, to obtain a rotation angle on the order of one radian even when the modulating magnetic field is not strong,
one may resort to bulk graphite, for instance, flexible graphite.

Here we study Faraday rotation from a different aspect.
We first construct a periodic structure using graphene sheets.
Assuming that the periodic length is far smaller than the wavelength of incident light in the structure, the periodic graphene structure (PGS) can be considered a medium.
In this paper we study Faraday rotation when an incident linearly polarized electromagnetic wave passes through the medium (or PGS).

In such a PGS a collective resonance mode, a plasmon excitation, has a role. While there are only two parameters in monolayer or multilayer graphene,\ct{nature} incident wave (angular) frequency $\omega$ and cyclotron frequency $\omega_c$, there is a third parameter in bulk graphene, plasmon frequency $\omega_p$, which complicates our analysis. However, we will show that one can still produce an MO device with a high VC.

\section{ANALYSIS OF FARADAY ROTATION EFFECT}

In semiclassical theory the VC is\ct{fre}
 \bee \label{class}
 V=\frac{e\mu}{2m_e c}\omega\frac{dn}{d\omega},
 \ee
where $c$, $\mu$, $e$, $m_e$, $n$ and $\omega$ are the velocity of light in vacuum,
permeability of the medium, charge, electron mass, refractive index of the
medium, and angular frequency of the incident light respectively. The equation tells
us that to increase the VC, one may choose strong dispersion materials,
for instance, MO glass. However, noting that $\frac{e}{m}$
(multiplied by $B_0$) is the cyclotron angular frequency, we have
another approach to increase the VC.

As shown by Eq. (\ref{class}), the strength of the MO
rotation effect in materials is determined by the difference in
velocity between the right circularly polarized and left circularly polarized
waves. This difference reflects the asymmetry of the carrier response to an
external magnetic field. In the medium,
the carrier response to the external magnetic field is described by
cyclotron motion, $\omega_c=\frac{eB_0}{m_{eff}}$, where  $B_0$ and
$m_{eff}$ are the external magnetic field and carrier effective mass
respectively. However, from the expression for $\omega_c$, one can also reduce carrier
effective mass to enhance the carrier response to the external magnetic field.

\begin{center}
\begin{minipage}{0.4\textwidth}
\centering
\includegraphics[width=2.2in]{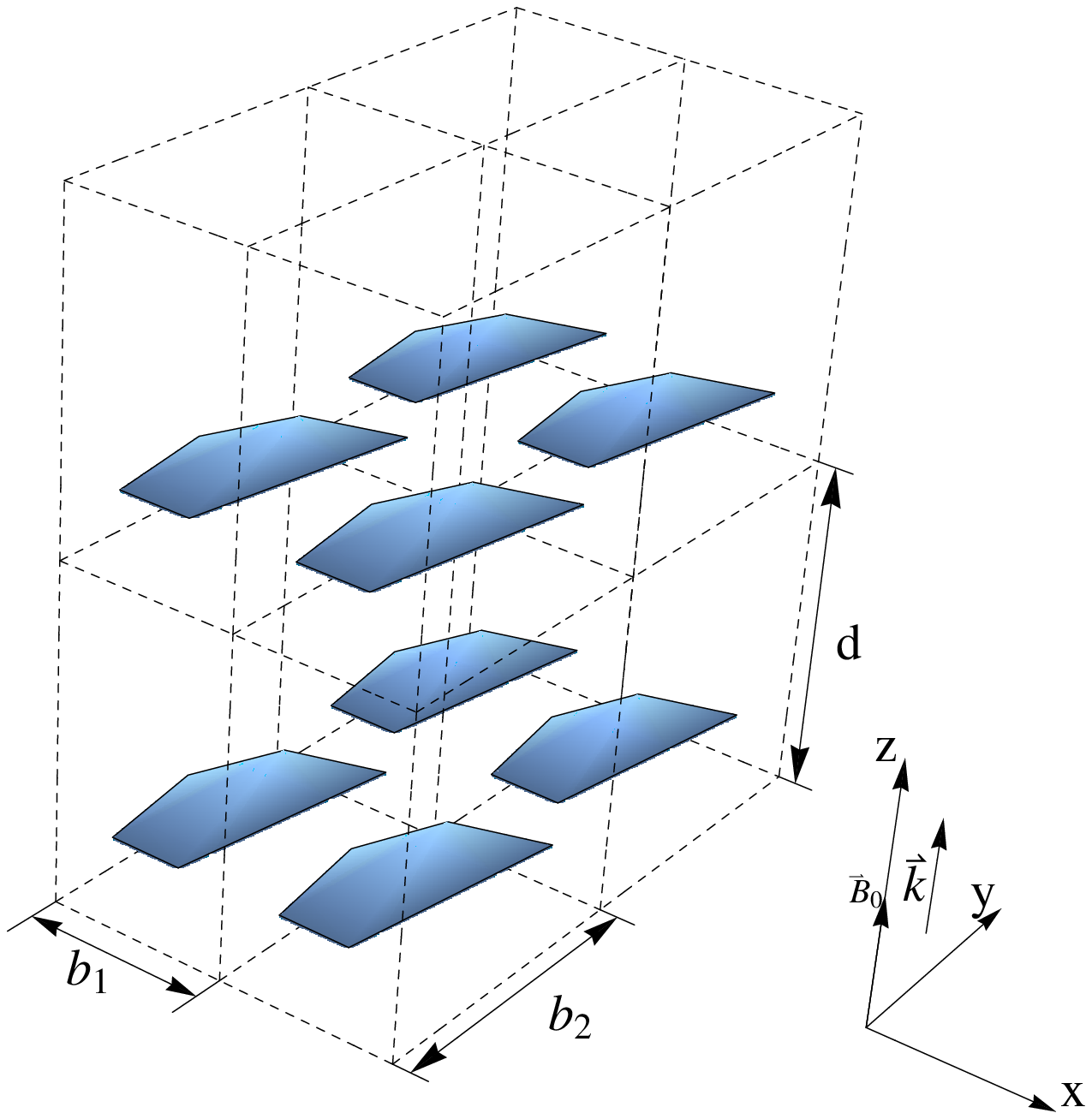}%
\figcaption{Periodic structure of graphene sheets. The sides of
the block cell are ${\mathbf{b}}_{1}$, ${\mathbf{b}}_{2}$ and
${\mathbf{d}}$, which are parallel to the $x$, $y$ and
$z$ axes, respectively. We impose the modulating magnetic induction
${\mathbf{B}_0}$ and propagation direction of the electromagnetic wave
both along $z$-axis.
  } \label{structure}
\end{minipage}
\setlength{\intextsep}{0.in plus 0in minus 0.1in} 
\end{center}

In usual materials $m_{eff}\sim m_e$, and such enhancement
is not significant. Notice the massless Dirac fermion
behavior of quasielectrons in graphene,\ct{discovery} {\it i.e.} $m_{eff}=0$, which means the
response to the external magnetic field should be sensitive. It is then natural to
use graphene to get a high VC. In this section we study the
Faraday MO effect in a PGS and
find that we can use this PGS to get a very high VC.

We first construct the PGS as in Fig. 1. The elementary cell for such a PGS is
defined as a rectangular block. The sides of the block are ${\mathbf
b}_1$, ${\mathbf b}_2$ and ${\mathbf{d}}$, which are parallel to the
$x$, $y$ and $z$ axes, respectively, and all of them are far lower
than the wavelength in the PGS. The
graphene sheet outspreads in the $x-y$ plane in each elementary cell.  The ratio of the
area of graphene to the area of a primitive cell in the $x-y$ plane,
$b_{1}b_{2}$, is $r$. Here we just consider the case of $r\equiv 1$,
that is, each graphene sheet completely covers the $x-y$ plane.

There are two types of fields in the structure (medium). One
is the modulating static magnetic field $\mathbf{B}_0$, and
the other is the electromagnetic wave field with $\mathbf{E}$ and
${\mathbf B}$. Here we assume that the modulating magnetic fields are far
stronger than the electromagnetic wave fields, {\it i.e.} $B_0\gg B$.
We impose the modulating magnetic induction $\mathbf{B}_0$ and
propagation direction of electromagnetic wave both along the $z$-axis.
Hence $\mathbf{B}$ and electric field ${\mathbf{E}}$ are in the $x-y$
plane.

Landau diamagnetism is
negligible because there is no external electric field. Therefore, we approximate the relative permeability
as $\mu_r\approx 1$.

Now we use the concept of an effective plasmon to study dielectric permittivity.
We first write the effective carrier density, $n=n_{2}r/d$,
where $n_{2}=\frac{g}{4\pi }(\frac{E_F}{\hbar v_{F}})^{2}$ is
the number of carriers per unit area, $E_F$ is the Fermi energy
(or chemical potential at zero temperature) without external fields, $v_F=1\times 10^6m/s$ is the
Fermi velocity of graphene and $g=4$ is the degeneracy.\ct{epl} In the graphene sheet, the Fermi energy $E_F$
and Fermi momentum $p_F$ satisfy $p_F=E_F/v_F=\hbar
\sqrt{\pi n_2}$.

Notice that since the Fermi velocity $v_{F}=\frac{c}{300}\ll c$,
the long-wave limit is always correct in
the periodic structure. If we turn off the modulating field, that is,
we ignore the cyclotron motion, the dielectric permittivity can be
written as a scalar,\ct{prl,prb,other}
\begin{equation}
\epsilon (\omega )=1-\frac{\omega _{p}^{2}}{\omega ^{2}},  \label{deg}
\end{equation}%
in the long-wave limit, where $\omega_p$ is the plasmon angular frequency or resonant angular frequency. We
emphasize that, as shown in Refs. \cite{prl}, \cite{prb} and \cite{apa}, this form
for dielectric permittivity is valid both in the quantum and classical theories, but the expression for $\omega_p$ in the quantum theory
is different from that in the classical. In the classical theory,
$\omega_p=\sqrt{\frac{ne^{2}}{m\epsilon _{0}}}$, while in graphene,\ct{prl,prb,other}
$\omega_p=\sqrt{\frac{ne^{2}v_F}{
p_F\epsilon _{0}}}$ due to quantum and 'relativistic' effects, where $\epsilon_0$ is the vacuum permittivity. In this case, the
permittivity is isotropic the in $x-y$ plane and there is no rotation effect.

Now turn on $B_0$ along the $z-$axis and discuss the rotation effect. To make semiclassical theory apply, we first
assume $\omega_c=\frac{eB_0v_F}{p_F
},\omega_p, \omega\ll\frac{E_F}{\hbar}$.\ct{note} Since $\hbar\omega \ll
E_F$, the photon energy is much lower than $E_F $, so we
consider only the cyclotron motion near the Fermi energy. This is
because the carriers with energy far lower than the Fermi surface
cannot absorb energy from a photon because of the Pauli exclusion
principle. Under the assumption, as pointed out by the Appleton-Hartree theory, we
write the relative dielectric permittivity $\epsilon_r$ in the $x-y$
plane as\ct{apa}
\begin{equation}
\epsilon_r =\left(
\begin{array}{cc}
\epsilon _{1} & \text{i}\epsilon _{2}  \\
-\text{i}\epsilon _{2} & \epsilon _{1}
\end{array}
\right),  \label{appl}
\end{equation}
where $\epsilon_1=1+\frac{\omega_p^2}{\omega_c^2-\omega
^2}$, $\epsilon_2=\frac{\omega_p^2\omega_c}{\omega
(\omega_c^2-\omega^2)}$,
$\omega_p=\sqrt{\frac{ne^{2}v_F}{p_F\epsilon _0}}$ and
$\omega_c=\frac{eB_0}{m_c}=\frac{eB_0v_F}{p_F}$. We have shown in
Ref. \cite{apa} that $\omega_p$ and $\omega_c$ are quantum
angular frequencies of the plasmon and cyclotron respectively. Equation
(\ref{appl}) tells us that the
effective permittivity is determined by two physical behaviors in the PGS,
namely, cyclotron motion and plasmon excitation,
characterized by two frequencies, $\omega _{c}$ and $\omega _{p}$
respectively. The cyclotron motion is determined by the modulating
magnetic field and the plasmon is a collective mode. On
one hand we find that, if $\omega _{c}\ll \omega,\omega _{p}$, the dielectric
permittivity tensor will degenerate into Eq. (\ref{deg}), while on
the other hand, $\epsilon _{2}\propto \omega _{c}/\omega $ will lead
to the optical rotation effect.

Maxwell's equations in the PGS are
\bee
\left\{
\begin{array}{rcl}
  \nabla\times \vv{E} &= &-\frac{\partial}{\partial t}\vv{B}, \\
  \nabla\times \vv{H} &=&  \frac{\partial}{\partial t}\vv{D},\\
  \nabla\cdot\vv{B} &= &0, \\
  \nabla\cdot\vv{E} &= &0,
\end{array}
       \right.
\ee
and constitutive relations are
\bee
\left\{
\begin{array}{rcl}
  \vv{B} &= &\mu_0\vv{H}, \\
  \vv{D} &=&  \epsilon_0\epsilon_r\cdot\vv{E},
\end{array}
       \right.
\ee
where $\mu_0$ is the vacuum magnetic permeability.

Supposing the electric field and magnetic field can be written as $
\vv{B}(\vv{E})=\vv{B}_0(\vv{E}_0)e^{\ci\omega t-\ci kz} $,
where $\vv{B},\,\vv{E},\,\vv{B}_0$ and $\vv{E}_0$ are vectors in
the $x-y$ plane, one finds that for the electric field
 \bee \frac{k^2
c^2}{\omega^2}\vv{E}_0-\epsilon_r\vv{E}_0=0,
 \ee
 where the identity $\mu_0\epsilon_0=1/c^2$ is used. The relation for
the magnetic field is similar. For each $\omega$, corresponding to solutions of
nonzero $\vv{E}=(E_x,\,E_y)$ in the above equation, there are two
eigenmodes: $k_1=\frac{\omega}{c}\sqrt{\epsilon_1+\epsilon_2}$ with
eigenvector $\vv{E}_{01}=(\hat{e}_x-\ci\hat{e}_y)A/2$ and
$k_2=\frac{\omega}{c}\sqrt{\epsilon_1-\epsilon_2}$ with eigenvector
$\vv{E}_{02}=(\hat{e}_x+\ci\hat{e}_y)A/2$, where $A$ is a constant. The
former eigenmode is left circularly polarized while the latter
eigenmode is right circularly polarized. Since $k_1\neq k_2$
for a certain $\omega$, the velocities of these two circularly
polarized waves are different. Therefore, a rotation
effect occurs for a linear electromagnetic wave passing through the
PGS; that is, if the incident wave is linearly polarized from the negative $z$-axis, the emitting wave is a still linearly polarized
but with a rotational polarized direction.

To analyze the rotation effect, we first suppose that both left and right circularly polarized waves can propagate in the medium and write out the electric fields of these waves as
\bee \left\{
\begin{array}{c}
  \vv{E}_1=\frac{A}{2}(\cos(\omega t-k_1z)\hat{e}_x+\sin(\omega t-k_1z)\hat{e}_y), \\
  \vv{E}_2=\frac{A}{2}(\cos(\omega t-k_2z)\hat{e}_x-\sin(\omega t-k_2z)\hat{e}_y),
\end{array}
\right.
\ee
with total $\vv{E}=\vv{E}_1+\vv{E}_2$.
This expression means that in the $z=0$ plane the linear polarization is along the $x-$axis.
We get
\beea
\vv{E}&=&A\cos(\omega t-\frac{k_1+k_2}{2}z)\cos(\frac{k_1-k_2}{2}z)\hat{e}_x \notag \\ &&
-A\cos(\omega t-\frac{k_1+k_2}{2}z)\sin(\frac{k_1-k_2}{2}z)\hat{e}_y
 \eea
Therefore, the total wave vector is
\bee
k=\frac{\omega}{2c}(\sqrt{\epsilon_1+\epsilon_2}+\sqrt{\epsilon_1-\epsilon_2}),
\ee
and the rotatory power is
\bee
\alpha=-\frac{\omega}{2c}(\sqrt{\epsilon_1-\epsilon_2}-\sqrt{\epsilon_1+\epsilon_2}).
\ee

To consider rotatory power in detail we introduce two
dimensionless quantities, $Q=\omega_c/\omega_p$ and
$\omega_0=\omega/\omega_p$. Then,
$\epsilon_1=1+\frac{1}{Q^2-\omega_0^2}$ and
$\epsilon_2=\frac{Q}{\omega_0(Q^2-\omega_0^2)}$. We find that
when $\omega_0\geq\frac{1}{2}(Q+\sqrt{Q^2+4})$, both
$\epsilon_1\pm \epsilon_2>0$, and a rotation effect occurs.
Notice that if we take $k<0$, the rotatory power is
still $\alpha$. This phenomenon is the well-known non-reciprocity of
the MO effect.

Now, we have
 \beea \frac{c\alpha}{\omega_p} &=&
 \frac{\omega_0}{2}\{\sqrt{1+\frac{1}{Q^2 -
\omega^2_0}+\frac{Q}{\omega_0(Q^2-
\omega_0^2)}} \nonumber \\ &&
-\sqrt{1+\frac{1}{Q^2 -
\omega^2_0}-\frac{Q}{\omega_0(Q^2- \omega_0^2)}}~\}.
 \eea
 In fig. 2(a) we show $\frac{c\alpha}{\omega_p}$ vs.
$\omega_0$ with different $Q$. We find that for fixed (high) $\omega_0$ and
$\omega_p$, $|\alpha|$ increases with decreasing $\omega_0$ or increasing Q (and thus $B_0$). The behavior agrees with that of multilayer graphene, as Fig. 3 in Ref. \cite{nature} shows. (The relative negative sign is due to the fact that we take $\epsilon_F>0$ while in Ref. \cite{nature} $\epsilon_F<0$).  At
$Q=\frac{\omega_0^2-1}{\omega_0}$, $\alpha$ reaches its maximum
value, $\frac{c\alpha}{\omega_p}=\omega_0\sqrt{\frac{\omega_0^2-1}{4\omega_0^2-2}}$, which can also be seen in Fig. 3. However, unlike single- and multilayer graphene, one cannot get a further step at this time. If we further reduce the incident frequency, one of the circularly polarized waves should decay in the medium, and the emitting wave should generally be elliptically polarized if the thickness of the medium is finite. Notice that for single and multilayer graphene, the decay is not important and we can also get a linearly polarized emitting wave.\ct{nature}

\begin{center}
\begin{minipage}{0.4\textwidth}
\centering
\includegraphics[width=2.4in]{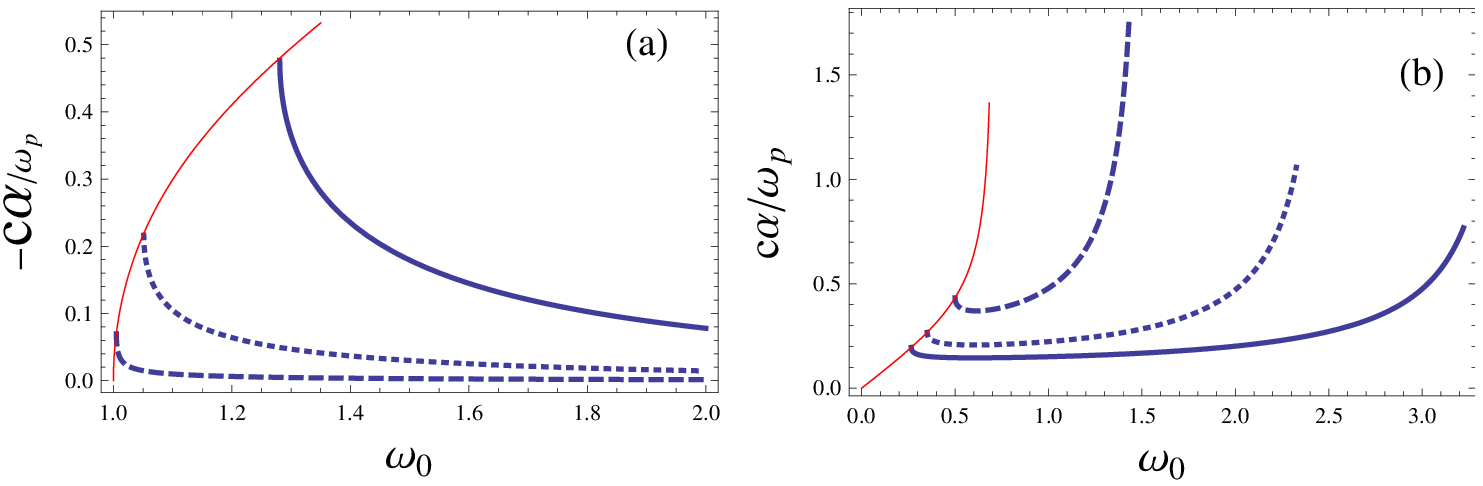}%
\figcaption{
$\frac{c\alpha}{\omega_p}$ vs. $\omega_0$ with different $Q$. (a) small $Q$ and higher $\omega_0$. (b) larger $Q$ and lower $\omega_0$. In figure (a) the
dashed line, dotted line and solid line are for $Q=0.01,\,0.1,\,$ and $0.5$ respectively and the red line is $\omega_0\sqrt{\frac{\omega_0^2 - 1}{4\omega_0^2 - 2}}$.
In figure (b) the
dashed line, dotted line and solid line are for $Q=1.5,\,2.5,\,$ and $3.5$ respectively and the red line is $\omega_0 \sqrt{\frac{1- \omega_0^2}{2- 4 \omega_0^2}}$.
  }
\end{minipage}
\setlength{\intextsep}{0.in plus 0in minus 0.in} 
\end{center}

Ordinarily, $Q$ is a small quantity. Theoretically, one may also study the case of large $Q$. One finds that if $Q>1/\sqrt{2}$ and $\frac{1}{2}(\sqrt{4+Q^2}-Q)<\omega_0<Q$, both right and left circularly polarized wave can pass the PGS without decay. We show the result of lower frequency with larger $Q$ at Fig. 2(b). Notice in this region rotatory power $\alpha>0$. However, in this region the lower $\omega_p$ and $\omega_c$ make the quasielectron behavior dominated by quantization,  thereby making our results doubtful.

 Figure 2(a) shows that $|\alpha|$ is on the order of
 $\omega_p/c$ when the frequency of the incident wave is close to the resonant frequency. For instance,
 when we adjust the parameters to $E_F=0.3\,eV$, $d=205\,\textrm{nm}$ and $Q=0.01$ ($B_0\simeq 4.7\textrm{T}$), we find that
 $\hbar\omega_p=0.09\textrm{eV}$ and $|\alpha|$ is on the order of
 $10-10^2\,\textrm{rad/mm}$, which is significant.

Fig. 3 shows that at small $Q$, rotatory power is proportional to
$Q$, or in other words, $\alpha\propto B_0$. From the Taylor
expansion we get the VC with
 \beea
 V&=&\frac{|\alpha|}{B_0/\mu_0}=\frac{e v_F^2\mu_0}{2c E_F\omega_0\sqrt{\omega_0^2-1}} \nonumber \\
 &=&\frac{5.73}{x(\mathrm{eV})\omega_0\sqrt{\omega_0^2-1}}(\textrm{min/Oe$\cdot$ cm}),
 \eea
where we assume $E_F=x(\textrm{eV})$. In the structure we generally choose
$x\sim 0.1\textrm{eV}$. When $Q<<1$, one way to improve $V$ is setting
The parameter to make frequency close to the resonant frequency,
$\omega_0\stackrel{>}{\sim} 1$. If we set
$\omega_0\sqrt{\omega_0^2-1}\sim 1$ and set $E_F=0.3eV$, $V$ is
on the order of $19\,\mathrm{min/Oe\cdot cm}$. The VC of the graphene
structure is about one or two orders larger than that of
MO glass. This is thus a significant advancement in
magnetic optics.

\begin{center}
\begin{minipage}{0.4\textwidth}
\centering
\includegraphics[width=2.0in]{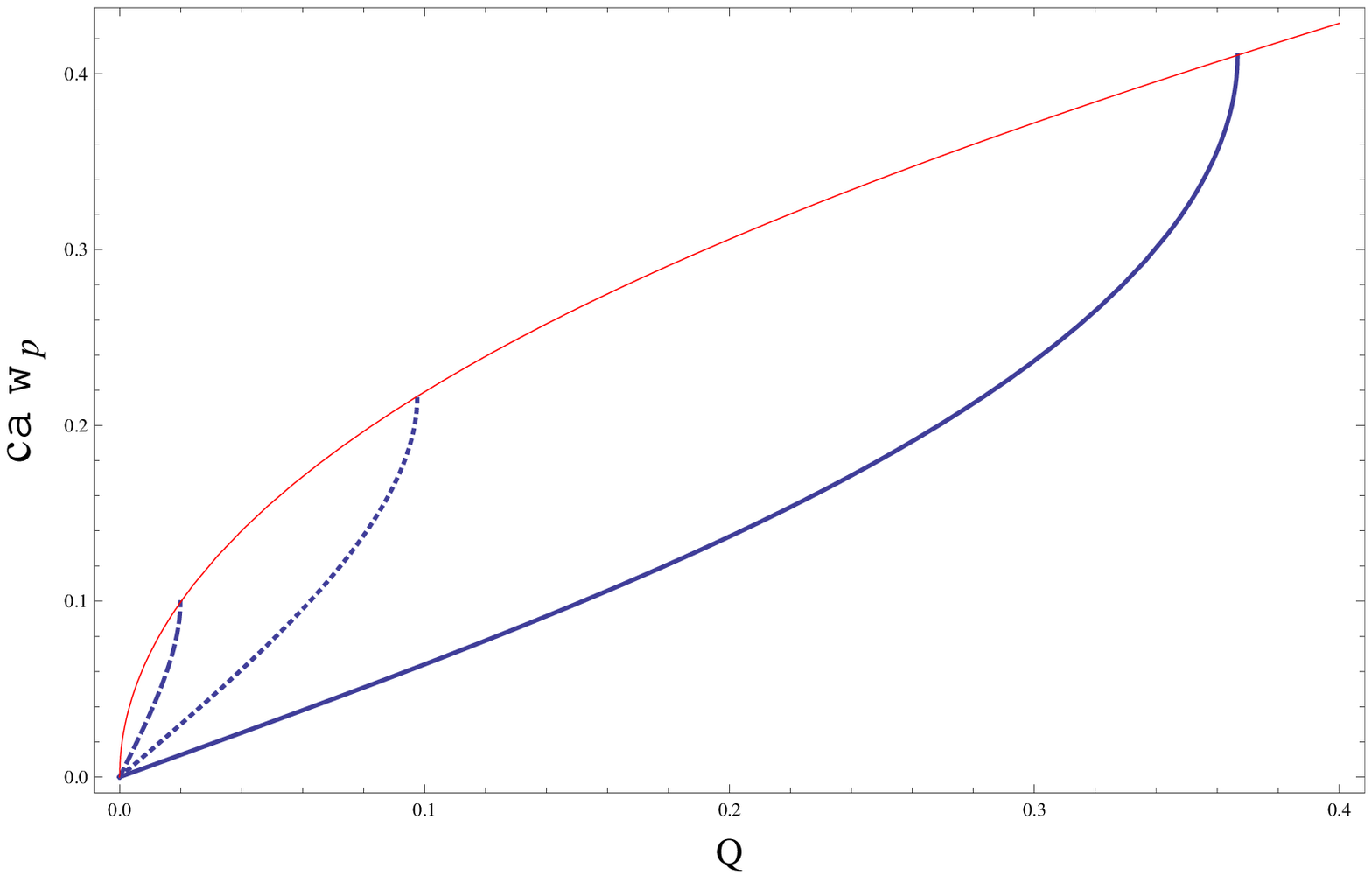}%
\figcaption{
$\frac{c\alpha}{\omega_p}$ vs. $Q$ with different $\omega_0$. The
dashed line, dotted line and solid line are for $\omega_0=1.01,\,1.05,\,and 1.2$ respectively. The red line is
$\frac{1}{2}(Q +\sqrt{4 + Q^2})\sqrt{\frac{Q}{3Q+\sqrt{4 +Q^2}}}$.
  }
\end{minipage}
\setlength{\intextsep}{0.in plus 0in minus 0.1in} 
\end{center}

Equation (\ref{class})
shows that $V$ in a usual material is
inversely proportional to $m_{eff}$, where $m_{eff}$ is the
effective carrier mass. In graphene, $m_{eff}=0$ and the
usual expression is not valid. However, $p_F/v_F$ plays the same
role as $m_{eff}$, and thus one can obtain a very large (not
infinite) VC.
Here we estimate the order of enhancement. In usual materials, the carrier mass
$m_{eff}\sim m_e\simeq 10^5\,eV/c^2$. However, in graphene, cyclotron motion mass plays the role of effective
mass, $m_c=E_F/v_F^2$. If we choose
$E_F=0.2eV$, we find that $m_c\sim 10^4eV/c^2\sim
10^{-1}-10^{-2} m_{e}$. Thus, the VC in the PGS is about one or two orders larger than in rare-earth-doped MO glass.


Furthermore, as shown in Fig. 3,  when $Q$ tends to
$\frac{\omega_0^2-1}{\omega_0}$, the rotatory power exhibits
nonlinear behavior and VC increases further.

\section{DISCUSSIONS}

In this manuscript we have discussed the MO rotation effect in a PGS. The masslessness of graphene carriers gives them a very sensitive response to an external field, and the VC of the PGS is very high. We predict that the VC in the structure is about one or two orders larger than in rare-earth-doped magneto-optical glass.

Furthermore, even if $\omega$ is not close to the resonant angular
frequency $\omega_p$, we may also improve $B$ to increase the VC, although
such improvement is constrained by the condition
$Q<\frac{\omega_0^2-1}{\omega_0}$.

In a usual material, the effective mass is always a constant,
whereas in graphene,
the role of $m_{eff}$,
$m_c=E_F/v_F^2$, is
adjustable. This behavior can further expand the application range
of the magneto-optical effect.

We conclude that besides MO glass, we can also use a graphene-sheet periodic structure to attain a high VC, increasing the VC up to one or two orders of magnitude. We hope our study is helpful in designing new types of MO devices.

\section*{ACKNOWLEDGMENTS}
The authors are very grateful to Prof. C.M. Zhang and Dr. M.G. Xia. This
work was supported by the Key grant Project of the Chinese Ministry of
Education (Grant No. 708082), the National Basic Research Program of
China (973 Program) (Grant No.2009CB623306), the International Science
\& Technology Cooperation Program of China (Grant No.2010DFR50480),
the National Nature Science foundation of China-NSAF (Grant No.
10976022), and the National Nature Science foundation of China
(Grants No. 50632030 and 11074196).

\balance

\end{multicols}


\begin{thebibliography}{99}

\bibitem{faraday}
M. Faraday, Philos. Trans.R.Soc. London \textbf{136}, 1 (1846); M. Faraday,
Philos.Mag. \textbf{28}, 294 (1846).

\bibitem{metry}
S. Pustelny, A. Wojciechowski, M. Gring, M. Kotyrba, J. Zachorowski and W. Gawlik, J.Appl.Phys. {\bf 103}, 063108 (2008).

\bibitem{high}
R.P. Tatam, M. Berwick, J.D.C. Jones, D.A. Jackson,
 Appl.Phys.Lett. {\bf 51}, 864 (1987).

\bibitem{nature}
I. Crassee, J. Levallois, A.L. Walter, M. Ostler, A. Bostwick, E. Rotenberg, T. Seyller, D. Marel and A.B. Kuzmenko, Nature Phys. 7, 48 (2010).

\bibitem{prb1}
A. Ferreira, J. Viana-Gomes, Y.V. Bludov, V. Pereira, N.M.R. Peres, and A.H. Castro Neto, Phys.Rev. {\bf B84}, 235410 (2011).

\bibitem{apl}
D. L. Sounas and C. Caloz, Appl. Phys. Lett. {\bf 98}, 021911 (2011).

\bibitem{prb2}
M. Orlita, {\it et al}, Phys. Rev. {\bf B83}, 125302 (2011).

\bibitem{prl2}
E. A. Henriksen, Z. Jiang, L.C. Tung, M. E. Schwartz, M. Takita,
Y.-J.Wang,P.Kim,and H.L.Stormer, Phys.Rev.Lett. {\bf 100}, 087403
(2008).

\bibitem{phil}
O. Roslyak, G. Gumbs and D. Huang, Phil. Trans. R. Soc. {\bf A368}, 5431  (2010).

\bibitem{fre}
C.Z. Tan and J. Arndt, J. Non-Cryst. Solids, {\bf 222}, 391 (1997).


\bibitem{discovery}
K.S. Novoselov, A.K. Geim, S.V. Morozov, D. Jiang, Y. Zhang, S.V. Dubonos, I.V. Grigorieva
and A.A. Firsov, Science, {\bf 306}, 669 (2004); K.S. Novoselov, A.K. Geim, S.V. Morozov, D. Jiang, M.I. Katsnelson, I.V. Grigorieva, S.V. Dubonos and
A.A. Firsov, Nature (London) {\bf 438}, 197 (2005); Y. Zhang, Y. Tan, Horst L. Stormer, and P. Kim, Nature (London) {\bf 438}, 201 (2005).

\bibitem{epl} D. Liu, S. Zhang, E. Zhang, N. Ma and H. Chen,
Europhys. Lett. \textbf{89} 37002 (2010).

\bibitem{prl} S.D. Sarma and E. H. Hwang, Phys.Rev.Lett. \textbf{102}, 206412 (2009).

\bibitem{prb} E. H. Hwang and S. Das Sarma, Phys.Rev. \textbf{B. 75}, 205418 (2007).

\bibitem{other} K.W.K. Shung, Phys.Rev. \textbf{B. 34}, 979 (1986);
T. Ando, J.Phys.Soc.Jpn. \textbf{75}, 074716 (2006); B. Wunsch, T.
Stauber, F. Sols and F. Guinea, New J.Phys. \textbf{8}, 318 (2006).



\bibitem{note}
Here we discuss briefly the constraint $\omega_c,\omega_p,\,\omega\ll E_F/\hbar$. Supposing $E_F=x\,eV$,
$\omega_c\ll E_F/\hbar$ means that $B_0\ll \frac{E_F^2}{e\hbar
v_F^2}\sim x^2\times 10^3T$, which is a very weak constraint. At the
same time, $\omega_p\ll E_F/\hbar$ means that $d\gg 5.75/x\,nm$,
which is still not a strong constraint. But because of the above
constraint for $d$, one cannot use common bulk graphite. To
satisfy the condition we may resort to flexible
graphite.\ct{graphite} Finally, $\omega\ll E_F/\hbar$ means that
$\omega\ll 1.5x\times 10^{15} s^{-1}$. Correspondingly, if we choose
$x=0.2$, then the constraints are $B_0\ll 40\,T$,
$d\gg 29\,nm$ and $\omega\ll 3\times 10^{14}\,s^{-1}$ respectively.

\bibitem{apa}
D. Liu, Z. Xu, N. Ma and S. Zhang, Appl. Phys. {\bf A106}, 945 (2012).

\bibitem{graphite}
D.S.L. Abergela, V. Apalkovb, J. Berashevicha, K. Zieglerc and
T. Chakraborty, Advances in Physics, {\bf 59}, 261 (2010).



\end{thebibliography}
\end{document}